\documentclass[twocolumn,aps,prl,showpacs,amsmath,amssymb,superscriptaddress,floatfix]{revtex4-2}
\usepackage[breaklinks=true,colorlinks,citecolor=blue,linkcolor=blue,urlcolor=blue]{hyperref}

\usepackage[T1]{fontenc}
\usepackage[latin9]{inputenc}
\usepackage{lineno}
\usepackage{textcomp}
\usepackage{amsmath}
\usepackage{graphicx}
\usepackage{multirow}
 \usepackage{makecell}
 \usepackage{booktabs}
\usepackage{bbold}
\usepackage{lipsum}
\usepackage{array}
\newcolumntype{P}[1]{>{\centering\arraybackslash}p{#1}}

\def\be{\begin{equation}}
\def\ee{\end{equation}}

\def\bea{\begin{eqnarray}}
\def\eea{\end{eqnarray}}

\makeatletter
\usepackage{lipsum}
\usepackage{epsfig}

\usepackage[colorlinks,linkcolor=blue]{hyperref}
\usepackage[nameinlink,capitalise]{cleveref}

\usepackage{subfigure}
\usepackage{color}
\usepackage{physics}

\definecolor{Red}{rgb}{1,0,0}
\definecolor{Blu}{rgb}{0,0,1}
\definecolor{Green}{rgb}{0,1,0}

\makeatother
\usepackage{amssymb}
\usepackage{babel}

\usepackage{tikz,xcolor,hyperref}
\definecolor{lime}{HTML}{A6CE39}
\DeclareRobustCommand{\orcidicon}{%
	\begin{tikzpicture}
	\draw[lime, fill=lime] (0,0)
	circle [radius=0.16]
	node[white] {{\fontfamily{qag}\selectfont \tiny ID}};
	\draw[white, fill=white] (-0.0625,0.095)
	circle [radius=0.007];
	\end{tikzpicture}
	\hspace{-2mm}
}

\foreach \x in {A, ..., Z}{%
	\expandafter\xdef\csname orcid\x\endcsname{\noexpand\href{https://orcid.org/\csname orcidauthor\x\endcsname}{\noexpand\orcidicon}}
}

\begin{document}

\title{Fast electrically switchable large gap quantum spin Hall states in MGe$_2$Z$_4$}

\author{Rajibul Islam}
\email{rislam@magtop.ifpan.edu.pl}
\affiliation{International Research Centre MagTop, Institute of Physics, Polish Academy of Sciences, Aleja Lotnik\'ow 32/46, PL-02668 Warsaw, Poland}
\author{Ghulam Hussain}
\affiliation{International Research Centre MagTop, Institute of Physics, Polish Academy of Sciences, Aleja Lotnik\'ow 32/46, PL-02668 Warsaw, Poland}

\author{Rahul Verma}
\affiliation{Department of Condensed Matter Physics and Materials Science,
Tata Institute of Fundamental Research, Colaba, Mumbai 400005, India}

\author{Mohammad Sadegh Talezadehlari }
\affiliation{Institute of Physics, Polish Academy of Sciences, Aleja Lotnik\'ow 32/46, PL-02668 Warsaw, Poland}
\affiliation{Institute of Physics, University of Rostock, Albert-Einstein-Straße 23-24, 18059 Rostock, Germany}

\author{Zahir Muhammad}
\affiliation{International Research Centre MagTop, Institute of Physics, Polish Academy of Sciences, Aleja Lotnik\'ow 32/46, PL-02668 Warsaw, Poland}
\affiliation{Hefei Innovation Research Institute, School of Microelectronics, Beihang University, Hefei 230013, P. R. China}

\author{Bahadur Singh}
\email{bahadur.singh@tifr.res.in}
\affiliation{Department of Condensed Matter Physics and Materials Science, Tata Institute of Fundamental Research, Colaba, Mumbai 400005, India}

\author{Carmine Autieri}
\email{autieri@magtop.ifpan.edu.pl}
\affiliation{International Research Centre MagTop, Institute of Physics, Polish Academy of Sciences, Aleja Lotnik\'ow 32/46, PL-02668 Warsaw, Poland}
\affiliation{Consiglio Nazionale delle Ricerche CNR-SPIN, UOS Salerno, I-84084 Fisciano (Salerno), Italy}

\begin{abstract}
	
	Spin-polarized conducting edge currents counterpropagate in quantum spin Hall (QSH) insulators and are protected against disorder-driven localizations by the time-reversal symmetry. Using these spin-currents for device applications require materials having large band gap and fast switchable QSH states. By means of in-depth first-principles calculations, we demonstrate the large band gap and fast switchable QSH state in a newly introduced two-dimensional (2D) material family with 1T$^\prime$-MGe$_2$Z$_4$ (M = Mo or W and Z = P or As). The thermodynamically stable 1T$^\prime$-MoGe$_2$Z$_4$ monolayers have a large energy gap around $\sim$237 meV. These materials undergo a phase transition from a QSH insulator to a trivial insulator with a Rashba-like spin splitting  under the influence of an out-of-plane electric field, demonstrating the tunability of the band gap and its band topology. Fast topological phase switching in a large gap 1T$^\prime$-MoGe$_2$Z$_4$ QSH insulators has potential applications in low-power devices, quantum computation, and quantum communication.  
	
\end{abstract}
\date{\today}
\maketitle

{\it Introduction:} The discovery of two-dimensional (2D) materials has opened a new era in condensed matter and materials physics owing to their conceptual novelties and potential for a wide range of device applications.~\cite{novoselov2005two,novoselov2004electric,zhang2005experimental,du2009fractional,novoselov2005two1,bhimanapati2015recent}. Main importance of the quantum spin Hall (QSH) insulators is the 1D counter-propagating conducting helical edge modes protected by the time-reversal symmetry inside the 2D bulk. These helical modes make QSH insulators appealing for dissipationless, fast, and energy-efficient device applications~\cite{PhysRevLett.95.146802,PhysRevLett.95.226801}. The existence of QSH states has been theoretically predicted in diverse families of 2D materials and quantum well structures through bulk and edge state computations\cite{Zhang21heterobilayer,Bernevig_QSH,Lau19}. Their experimental verification has been reported in 2D quantum wells of HgTe/CdTe and InAs/GaSb by measuring the quantized spin Hall conductance~\cite{bernevig2006quantum,konig2007quantum,QSH_InAs, QSH_InAs2}.

On realizing the QSH state in single-phase materials, the thin films of WTe$_2$ have been extensively studied over the last few years~\cite{Ok19}. The experimental signature of the QSH effect has been demonstrated recently in the monolayer films of the 1T$^\prime$ phase of WTe$_2$~\cite{Fei2017, Zheng2016, Wu18, PhysRevX.11.041034}. 
In van der Waals and 2D materials, many QSH candidates have proposed and investigated\cite{Lodge2021,Kim2021}. However, for both quantum wells and 2D materials, the longest topological protection length is of the order of few micrometers\cite{Wu18,Lunczer2019}. It has been proposed that different mechanisms may contribute to the shortening of the topological protection length such as charge puddles\cite{Vayrynen2013}, spontaneous time-reversal symmetry breaking\cite{Pikulin2014,Paul2022} and acceptor dopants\cite{Dietl2022}. However, some of these mechanisms are valid just for the quantum wells\cite{Nguyen2022}.  
Considering also that the QSH phase in thin films and 2D materials can be tuned by electric field, strain\cite{Iolanda2021} and adatoms\cite{Zhang2018}, it is highly desirable to continue the search for new 2D materials with larger inverted band gaps and superior transport properties to achieve the room temperature QSH phase and engineering device applications.
\begin{figure}[b]
	\begin{center}
		\includegraphics[width=0.95\linewidth]{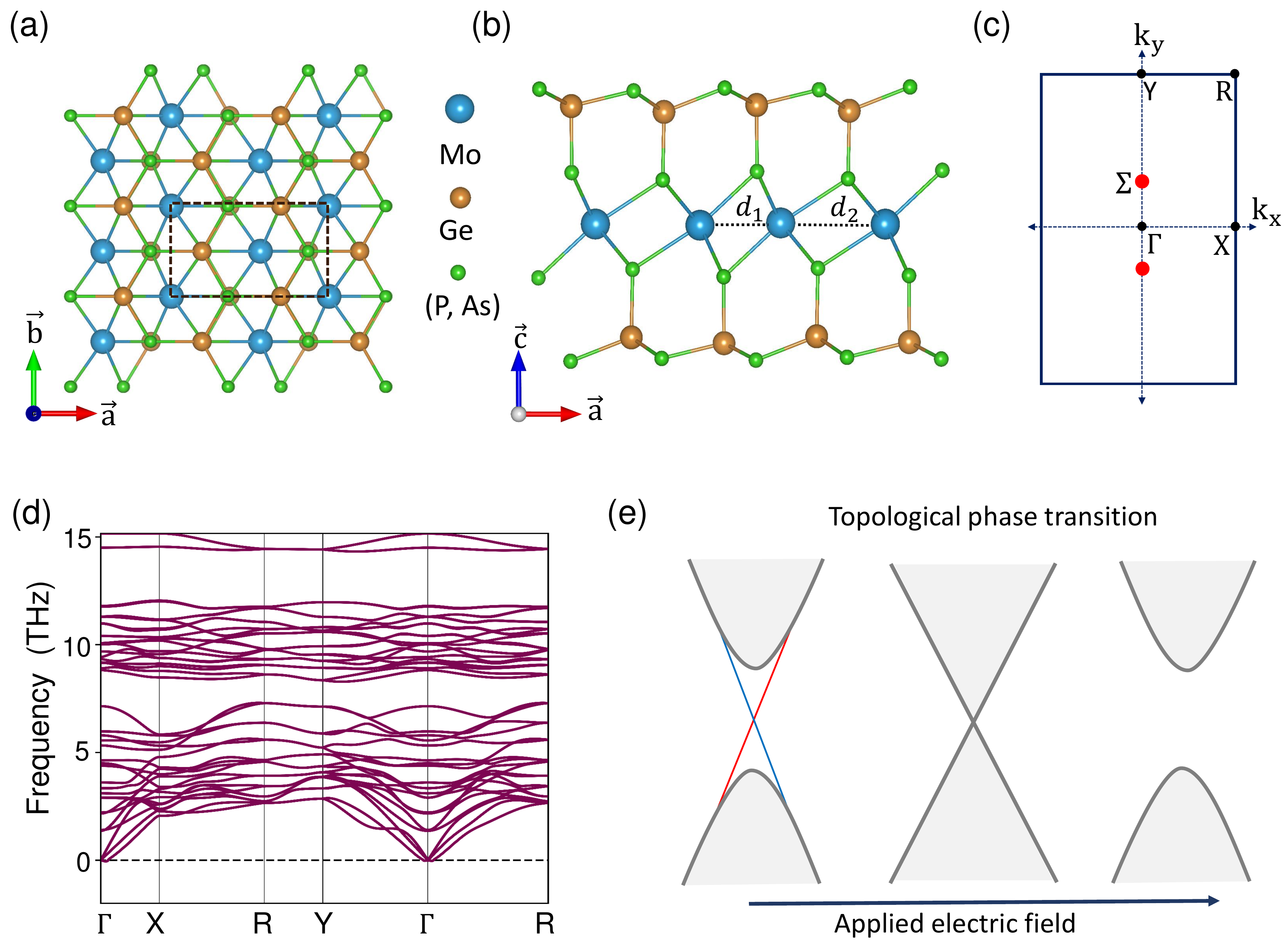}
	\end{center}
	\centering
	\caption{(a) Top view and (b) side view of the 1T'-phase of MGe$_2$Z$_4$ materials. (c) 2D Brillouin Zone (BZ) with high-symmetry points. At the red points $\Sigma$, the band structure without SOC closes the gap. (d) Phononic band structure of MoGe$_2$P$_4$ in the 1T' structural phase. (e) Schematic representation of the topological transition in external electric field.}
	\label{crystal_structure}
\end{figure}

Recently, a new MoSi$_2$N$_4$ class of 2D materials have been synthesized using a novel bottom-up technique~\cite{hong2020chemical,Yin2022emerging}. These materials show remarkable properties such as stability under ambient conditions, a semiconducting behavior, high mobility, and correlation-driven quantum anomalous Hall states, among other properties which are superior to the well-known transition metal dichalcogenides (TMDs) class of materials~\cite{hong2020chemical,bertolazzi2011stretching,cai2014polarity,yang2021valley, IslamMoSi2N4,li2020valley, KONG2022100814,feng2021valley}.  The structural phases of these materials are further amenable to realize a semiconductor-to-metal transition by defects, and strain engineering as well as by applying electric field~\cite{wu2021semiconductor, Zhang21heterobilayer}. The commonly investigated 2H phase of these materials is well characterized in connection to the formation of superlattices or Janus phases~\cite{Hussain22exploring, Hussain22strain, Hussain22emergence}.  More recently, we have shown that  MoSi$_2$N$_4$ materials can form a 1T$^\prime$ polytypic structure that realizes robust QSH insulator states with excellent transport properties~\cite{Islam_QSH_Si}. Motivated by these studies, we here introduce the new members in this family of QSH insulators that are thermodynamically stable and realize fast electric-field switchable helical edge states with excellent spin transport and properties. Based on our in-depth first-principles calculations, we show that the 1T$^\prime$-phase of MGe$_2$Z$_4$ (M = Mo or W, Z = P or As) monolayers have a large band gap of 237 meV. To our knowledge, this topological band gap is the highest reported in 1T$^\prime$-phases and comparable with the gap of 260 meV measured in the high-temperature quantum spin Hall of a higher-order topological insulator\cite{Shumiya2022}. We also report a controlled transition from the topological state to the trivial state under an external electric field.\\

{\it Methods:} Electronic structure calculations were performed within the framework of density functional theory using the projector augmented wave method using the VASP package~\cite{VASP}. The relativistic effects were considered self-consistently. A plane-wave energy cut-off of 500 eV was used. We used optPBE-vdW functional to consider exchange-correlation effects~\cite{perdew1996generalized}. Both the atomic positions and lattice parameters are optimized until the residual forces on each atom were less than 0.001 eV/{\AA} and the total energy was converged to 10$^{-8}$ eV with a Gaussian-smearing method.  For a more accurate estimation of the band order and the band gap, we employed HSE06 hybrid functional with 25\% of the exact exchange. We performed the calculations using a 6$\times$12$\times$1 $\Gamma-$centered \textbf{k}-mesh with 72  \textbf{k}-points in the Brillouin zone. The thermodynamical stability has been checked within the density functional perturbation theory framework using PHONOPY\cite{TOGO20151} code and the {\it ab initio} molecular dynamics simulations~\cite{MD} were performed using 2$\times$4$\times$1 supercell. We extracted the real space tight-binding Hamiltonian for calculated topological properties using the Wanniertools package~\cite{wu2018wanniertools}. The Wannier functions were generated using M $d$, Ge $s$ and $p$, and  Z $p$ orbitals within VASP2WANNIER~\cite{mostofi2008wannier90}. The surface energy spectrum was obtained within the iterative Green's function method~\cite{Greenwanniertools}.
\\
We obtain the SHC $\sigma_{xy}^z$ using the Kubo formula\cite{PhysRevLett.117.146403,PhysRevB.94.085410,PhysRevB.105.165140,PhysRevB.99.060408,PhysRevB.98.214402}-
\begin{equation}
	\sigma_{xy}^z = \frac{-e^2}{VN_\textbf{k}^3\hbar}\sum_\textbf{k} \Omega_{xy}^z(\textbf{k})
	\label{eq1}    
\end{equation}
where, 
\begin{equation}
	\Omega_{xy}^z(\textbf{k}) = \sum_n f(E_{n,\textbf{k}}) \Omega_{n,xy}^z(\textbf{k})
	\label{eq2} 
\end{equation}
is the \textbf{k}-resolved spin Berry curvature and
\begin{equation}
	\Omega_{n,xy}^z(\textbf{k}) = \hbar^2 \sum_{m\neq n} \frac{-2Im\bra{n\textbf{k}}j_x^z\ket{m\textbf{k}} \bra{m\textbf{k}}\hat{v}_y\ket{n\textbf{k}}}{(E_{n,\textbf{k}}-E_{m,\textbf{k}})^2}
	\label{eq3} 
\end{equation}
is the \textbf{k}- and band-resolved spin Berry curvature.
In Eqs. from (\ref{eq1}) to (\ref{eq3}), N$_\textbf{k}$ is the number of \textbf{k}-points in the BZ and $\bra{n\textbf{k}}$ denotes the Bloch state with energy E$_{n,\textbf{k}}$ and occupation $f(E_{n,\textbf{k}})$. The SHC $\sigma_{xy}^z$ represents the spin-current along the x-direction generated by the electric field along the y-direction, and the spin current is polarized along the z-direction. We have used a dense \textbf{k}-grid of 100 $\times$ 100$\times$ 100 for the spin Berry curvature and SHC, therefore N$_\textbf{k}$=100.\\

\begin{table*}[t]
	\caption{The calculated  structural parameters and elctronic properties of the 1T' phase of the MGe$_2$Z$_4$ (M = Mo, W, and Z = P, As) family are demonstated in this table. The lattice constants $a$ and $b$ and the bond lengths of the transition metal chain $d_1$ and $d_2$ are the structural parameters.  In the electronic properties, we report the band gap with   vdW (E$_g^{vdW}$) and HSE (E$_g^{HSE}$) exchange-functionals, as well as the inverted gap at the  $\Gamma$  point with vdW ($\delta^{vdW}$) and HSE ($\delta^{HSE}$) exchange-functionals. Lastly, we report the members of the family's topological invariant and topological phase.}
	
	\begin{tabular}{c |c |c| c| c| c|c|c|c}
		\hline \hline 
		Materials  &  a (\r{A})  &  b (\r{A}) & $d_1$ (\r{A}) &  $d_2$ (\r{A})  &  Band gap (meV)  &   Inverted gap (meV)  & Topological  & Topological \\
		&  	&     &      &     	& E$_g^{vdW}$  \hspace{0.5cm} E$_g^{HSE}$ & $\delta^{vdW}$  \hspace{0.5cm} $\delta^{HSE}$  & invariant & phase \\  
		\hline
		MoGe$_2$P$_4$    	&   6.382 		& 3.582      	& 3.040        	  &  4.326        	& 33.3 \hspace{0.5cm} 93.8      		& 503.7 \hspace{0.5cm} 877.0 	& $\mathbb Z_2 =1$ &  QSH \\
		MoGe$_2$As$_4$    	&   6.635 	& 3.738      	& 3.113        	  &  4.561        	& 45.9 \hspace{0.5cm} 136.9      		& 394.0 \hspace{0.5cm} 750.5 	& $\mathbb Z_2 =1$ &  QSH \\
		WGe$_2$P$_4$    	& 6.356		&  3.590      	& 3.037         	  & 4.310        	&  83.4 \hspace{0.5cm} 221.2    		& 604.7 \hspace{0.5cm} 951.2 & $\mathbb Z_2 =1$ &  QSH \\
		WGe$_2$As$_4$	&  6.600  &  3.7460  &  3.094           & 4.556     &  78.2 \hspace{0.5cm} 237.0     		& 494.9 \hspace{0.5cm} 912.9 & $\mathbb Z_2 =1$ & QSH	\\
		\hline  
		\hline
	\end{tabular} \label{T1:bulk}
\end{table*}

\begin{figure}[!b]
	\begin{center}
		\includegraphics[width=0.95\linewidth]{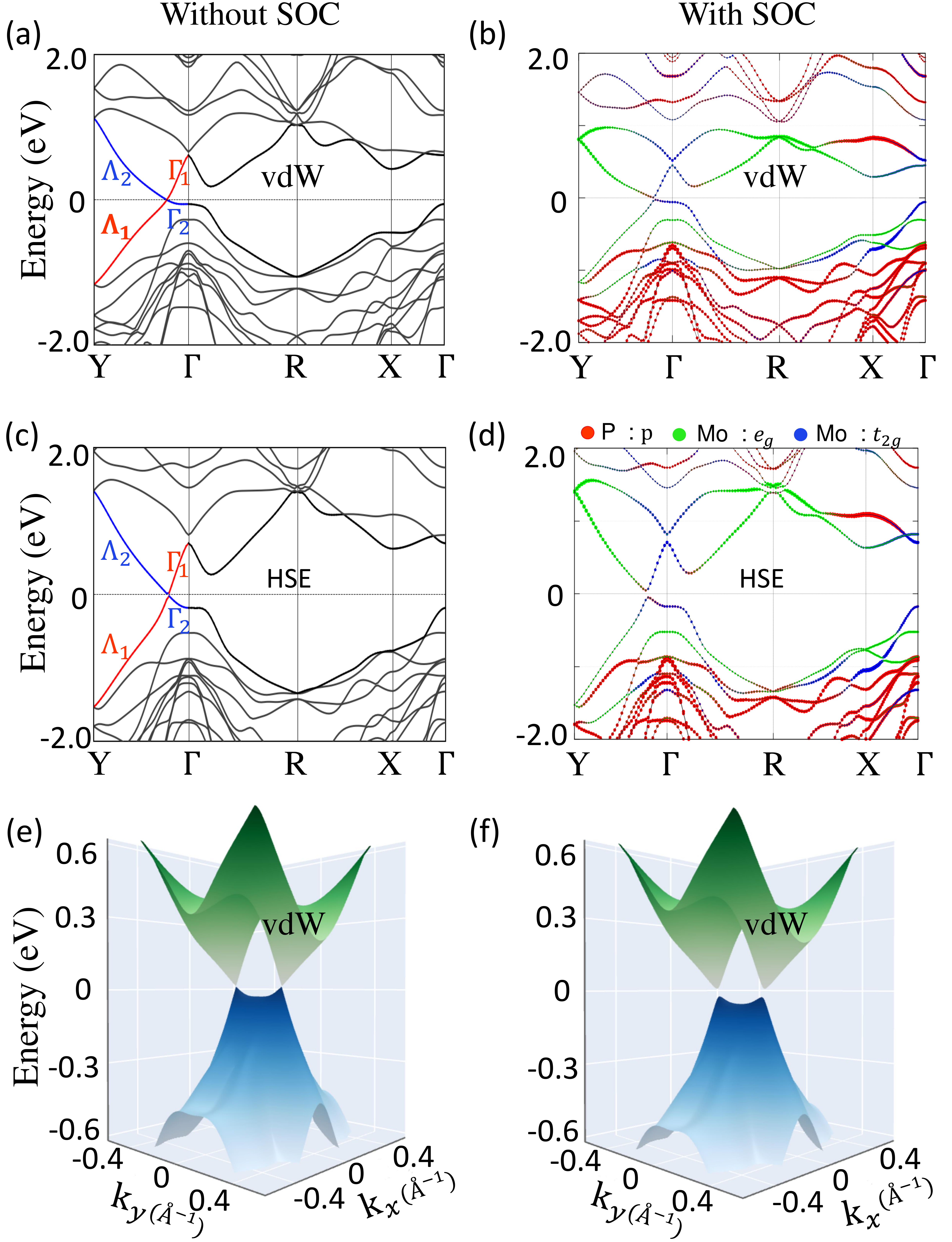}
	\end{center}
	\centering
	\caption{Electronic band structure of MoGe$_2$P$_4$ in the 1T'-phase  (a) without and (b) with SOC using the GGA functional with the vdW correction. and (c) without and (d) with SOC using the HSE functional. 2D band structures (e) without and (f) with SOC using the GGA functional with the vdW correction.  The Fermi level is set to zero.}
	\label{electronic_structure}
\end{figure}

{\it Crystal structure and symmetries:}
The crystal structure of the 2H-phase MGe$_2$Z$_4$  can be viewed as an MZ$_2$ layer sandwiched in between two GeZ layers, where the M atoms are located at the center of the trigonal prism building-block with six Z neighbours and the MZ$_2$ layer bonded vertically with the GeZ layers \cite{hong2020chemical}. This crystal structure has a hexagonal primitive cell with space group D$_{3h}^1$ (P$\Bar{6}$m2 No. 187). In the case of 1T'-phase MGe$_2$Z$_4$, the three atomic layers are stuck in such a way that the position of the M atom is at the center of  trigonal prism  building block twisted  by 60$^o$ and with six Z neighbours, which creates the octahedral building block of  M atoms with the six Z atoms in the MZ$_2$ layer. The different M-M bond lengths of atomic chains in the MZ$_2$ layer lowers the symmetry of the space group to p2$_1$/m$_1$ (No. 11) to form a rectangular primitive unit cell as shown in Fig.~\ref{crystal_structure}(a)-(b). Interestingly, the 1T' structure restores the inversion symmetry (I) in contrast to the 2H structure. The structural parameters are given in Table \ref{T1:bulk}.~\cref{crystal_structure}(c) shows the 2D Brillouin zone (BZ) of the 1T'-phase, the red dots reprensent band crossing points $\Sigma$ in the bandstructure without SOC. The absence of imaginary mode in the phonon band structure throughout the extended BZ in ~\cref{crystal_structure}(d) indicates the thermodynamical stability of the 1T'-phase of MoGe$_2$P$_4$.  Experimentally, the 1T'-phase of WTe$_2$ was synthesized successfully in past works. However, comparing to the structural stability of TMDs, we expect that the 1T'-phase of the MGe$_2$Z$_4$ compounds could be synthesized within certain thermal and mechanical conditions. Moreover, the topological phase transition in MGe$_2$Z$_4$ family under the influence of  external electric field is illustrated schematically in ~\cref{crystal_structure}(e). 

\begin{figure}[h]
	\includegraphics[width=0.99\linewidth]{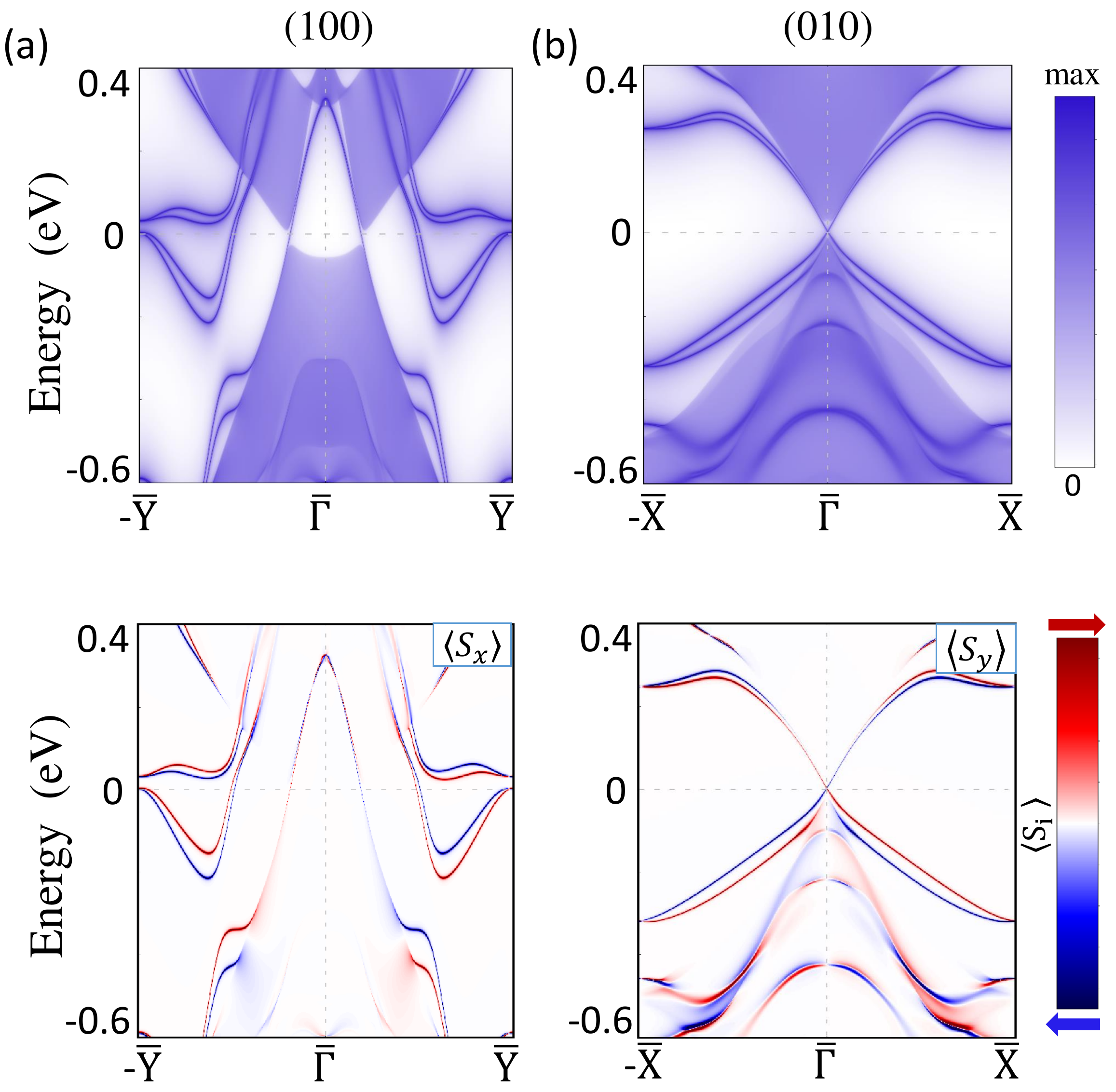}
	\caption{Top panels. Electronic band structure projected on the (a) (100) and (b)(010) edges of 1T'-MoGe$_2$P$_4$ within the GGA functional with the vdW correction. Bottom panels. The corresponding spin-polarized edge states are shown in red and blue for spin-up and spin-down, respectively.  The zero line marks the Fermi level.}
	\label{surface_state}
\end{figure}

{\it Electronic properties of MoGe$_2$Z$_4$ materials:} Now we will focus on the electronic band structure of 1T$^\prime$-MoGe$_2$Z$_4$ material. Fig.~\ref{electronic_structure}(a) shows the band structure of 1T$^\prime$-MoGe$_2$P$_4$ using the vdW correction without SOC. The valence and conduction bands cross each other along the $\Gamma$-Y line which indicates an inverted band ordering at the $\Gamma$ point. This band-crossing is protected, because the two bands belong to different symmetry operation ($\Lambda_1$ and $\Lambda_2$) representations. Additionally, a camel-back like feature has been observed along the $\Gamma$-Y line, which is typical of the system with large SOC and broken inversion symmetry as HgTe-based systems\cite{islam2021topological,PhysRevB.103.115209}. When the SOC is included, the band crossing is lifted and the system shows a band gap E$_g^{vdW}$ of 33 meV with the valence band maxima (VCM) and the conduction band minima (CBM) located at $\pm$(0, 0.102, 0){\AA}, respectively, as shown in Fig.~\ref{electronic_structure}(b). We have analysed the orbital character of the bands near the Fermi level, the  bands are dominated by the p-orbitals of P atoms and  d-orbitals of Mo atoms as in TMDs, the band inversion occurs between p of P and d of Mo at the $\Gamma$ point with an inverted gap ($\delta^{vdW}$) of ~503 meV. We have checked the robustness of our results with the more accurate HSE functional as shown in Fig.~\ref{electronic_structure}(c) without SOC and Fig.~\ref{electronic_structure}(d) with SOC. The HSE band structures show similar results to that obtained within GGA with vdW correction. The band gap (E$_g^{HSE}$) increases to ~93.8 meV while the inverted gap ($\delta^{vdW}$) increases to ~877 meV. Moreover, the 2D Wannier band structure also confirms the previous results [see Figs. ~\ref{electronic_structure}(e,f)] of a gapless Dirac semimetal phase in the absence of SOC and a QSH insulating phase in presence of SOC. The detailed data of the band gap and the inverted band gap of 1T$^\prime$-MoGe$_2$Z$_4$ materials are reported in Table ~\ref{T1:bulk}. The band structures of other materials are shown in the Supplementary Materials\cite{suppmat}.
\\
{\it Topological edge states:} One of the most important signature of QSH systems is the counter-propagating helical edge states. The electronic band structures with edge states projected on the (100) and (010) surface are shown in Fig. ~\ref{surface_state}(a) and (b), respectively. The topologically protected helical edge states connecting conduction and valence band are observed in the band structure projected on (010) edge. The associated spin-projected edge band structures are shown in the bottom panels. The different spin channels are indicated with different colours i.e spin-up in red and spin-down in blue, respectively. The 1T$^\prime$-MoGe$_2$P$_4$ is a QSH insulator similar to 1T$^\prime$ phase of the MSi$_2$Z$_4$ materials. Furthermore, we have calculated the Z$_2$ invariant with the Wannier charge centre (WCC) founding Z$_2$=1. The detailed study of the other members of the family is described in the Supplementary Materials\cite{suppmat}.\\    

\begin{figure}[b]
	\includegraphics[width=0.99\linewidth]{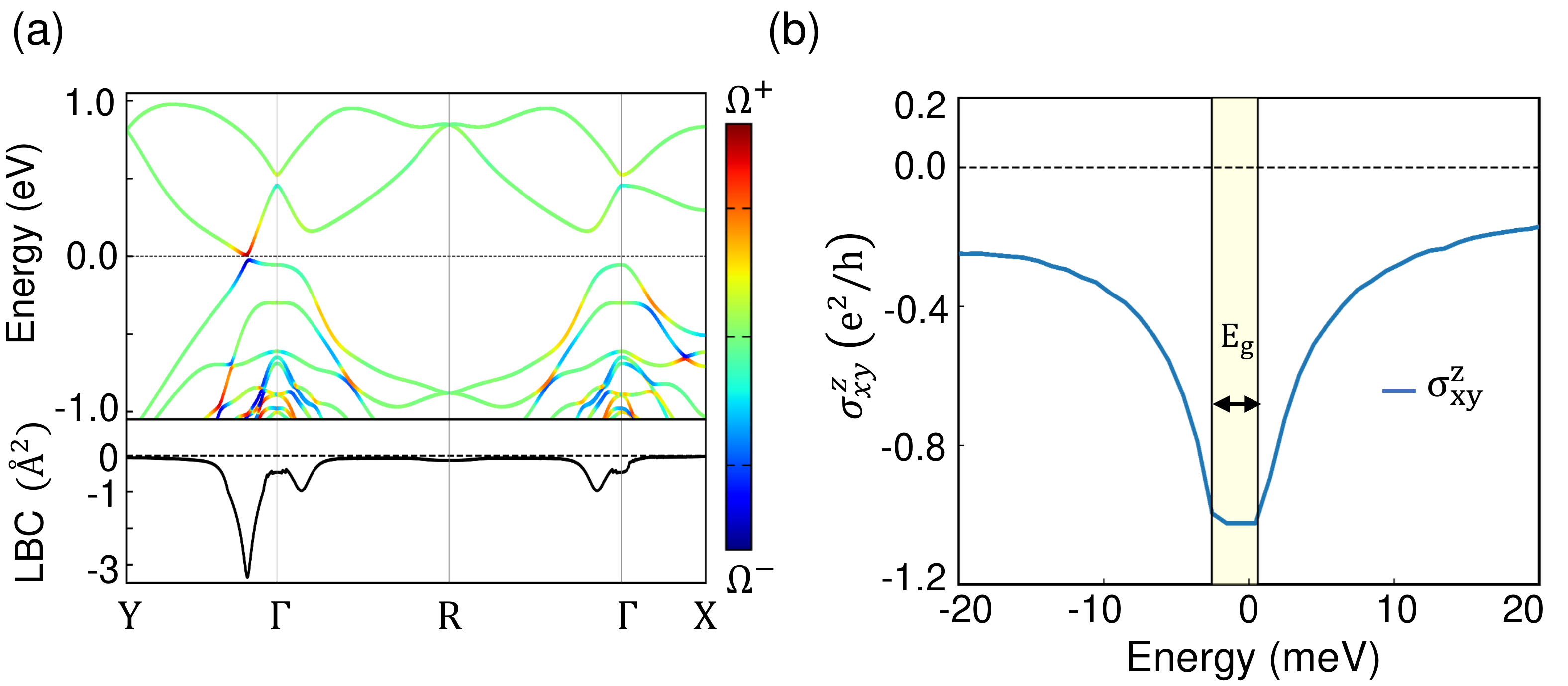}
	\caption{(a) Top panel. k-resolved spin Berry curvature (SBC) projected on  the electronic band structure. It can be seen that SBC is maximum at the avoiding crossing at the Fermi level along the ${\Gamma}Y$ direction. $\Omega^+$ and $\Omega^-$ represent the maximum and the minimum of the SBC, respectively. Bottom panel \textbf{k}-resolved spin Berry curvature in logarithmic scale (LBC). (b) Intrinsic spin Hall conductivity in units of $\frac{e^2}{h}$ for 1T'-MoGe$_2$P$_4$. The Fermi level is set to zero.}
	\label{Spin_Hall_Conductivity}
\end{figure}

{\it Spin Hall conductivity:}
We turn our discussion towards the spin Hall conductivity (SHC) and spin Berry curvature distribution. ~\cref{Spin_Hall_Conductivity}(a)  shows the distribution of the \textbf{k}-resolved spin Berry curvature (SBC) in the band structure. The SOC creates anticrossing gaps along ${\Gamma}Y$ and ${\Gamma}(-Y)$, therefore, a large opposite SBC arises on both sides of the anticrossing gaps. As the spin Hall conductivity is directly proportional to the integration of the SBC \cite{QSC_berry}, it leads to a large SHC of the $\sigma_{xy}^z$  as shown in ~\cref{Spin_Hall_Conductivity}(b), the value of SHC is quantized in the gap to a value of 1.0$\frac{e^2}{h}$, while for the other compounds is between 1.3$\frac{e^2}{h}$ and 1.6$\frac{e^2}{h}$ (See Supplementary Materials). The ideal value of the quantized SHC is 2$\frac{e^2}{h}$, however, deviations can occur due to the non-conservation of the z-component of the spin angular momentum, thanks to effects such as large spin-orbit, crystal field and hybridization\cite{Costa2021}.

\begin{figure}[t]
	\includegraphics[width=0.95\linewidth]{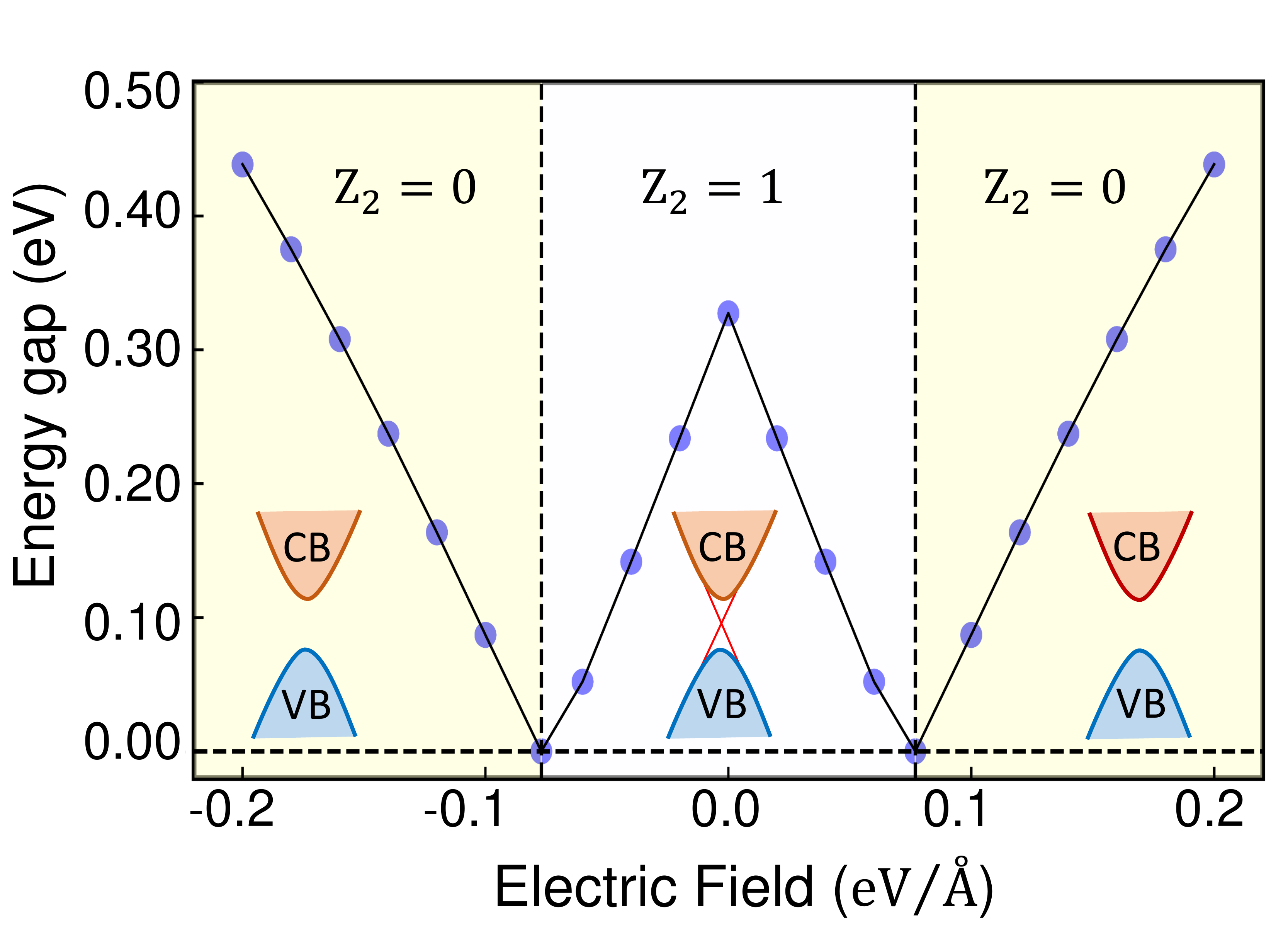}
	\vspace{-0.2cm}
	\caption{Topological nature (Z$_2$)  and  band gap of 1T$^\prime$-MoGe$_2$P$_4$  as a function of the applied out-of-plane electric field. The critical electric force fields qE$_C$=$\pm$0.077 eV/{\AA} are marked with vertical dashed lines..}
	\label{gapwithelectricfield}
\end{figure}

\begin{figure}
	\includegraphics[width=0.99\linewidth]{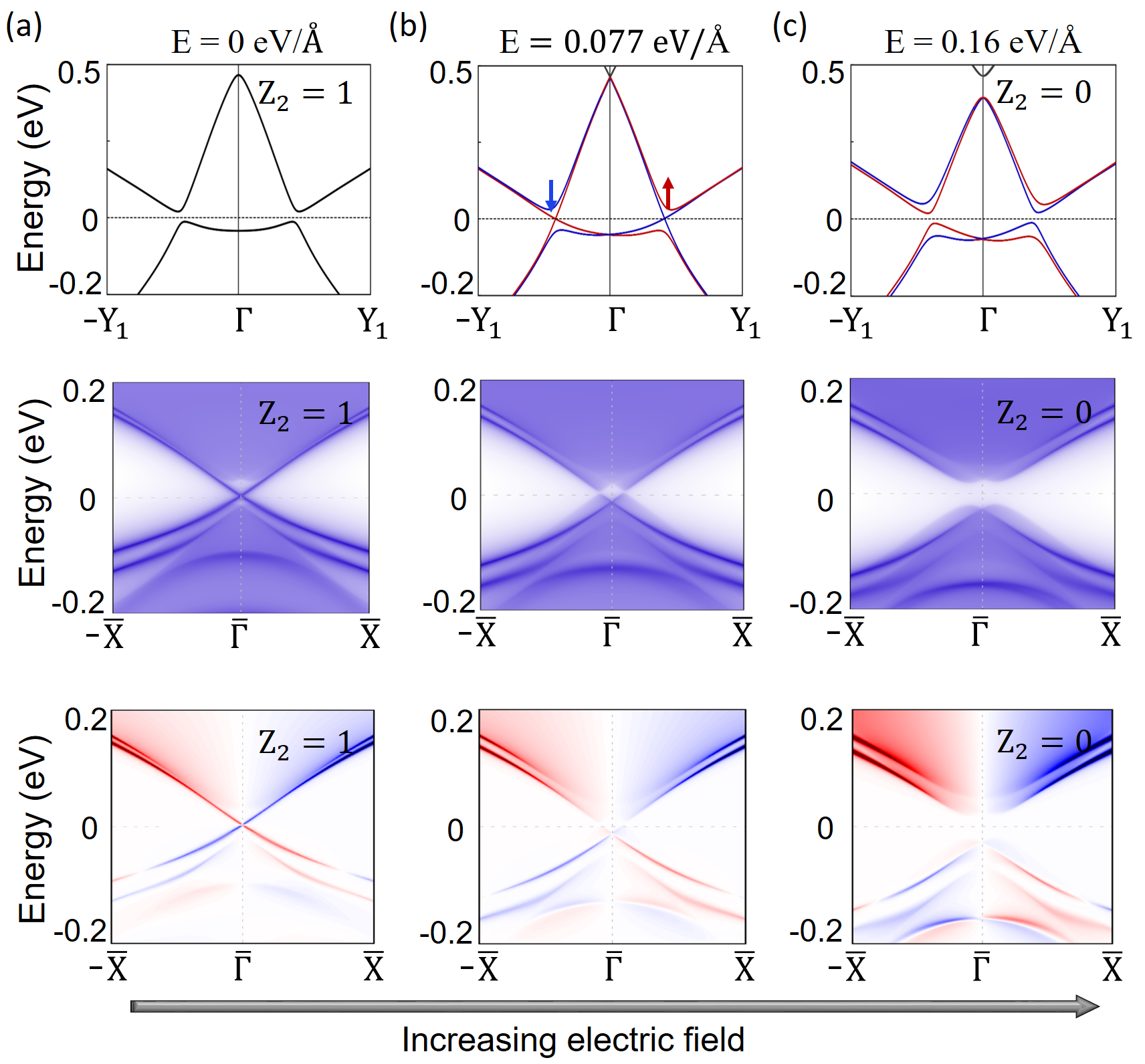}
	\caption{Electronic band structure band of the 1T'-MoGe$_2$P$_4$ monolayer at different values of the out-of-plane electric force field qE equal to (a) 0, (b) 0.077 and (c) 0.16 eV/{\AA}. The corresponding helical edge states with spin polarization for the (010) edge are shown in the middle and bottom panels, respectively.}
	\label{OUTPLANE_electricfield}
\end{figure}

{\it Electric field induced topological phase transition:}  In this subsection, we show the appearance of a phase transition induced by an external out-of-plane electric field. Increasing the vertical electric field (E), the band gap starts to reduce and reaches zero at a critical electric force field (qE$_C$= $\pm$0.077) eV/{\AA}, further increases in the electric field reopens the trivial gap. The topological phase transition from non-trivial ($\mathbb Z_2$=1) to trivial ($\mathbb Z_2$=0) via the application of out-of-plane electric field E is demonstrated in Fig.-~\ref{gapwithelectricfield}. A similar trend can be observed when the polarity of the electric field is reversed. The vertical electric field on 1T$^\prime$-MGe$_2$Z$_4$ compounds produces a charge imbalance between the GeZ sides, which changes the on-site potentials. Fig.~\ref{OUTPLANE_electricfield}(a) shows the band structure in the absence of the electric field, the bands are double degenerate as the system preserves both time-reversal symmetry and inversion symmetry. On the other hand, when the electric field is applied, the degeneracy of the bands is lifted showing a strong Rashba spin-splitting around the band gap due to the broken inversion symmetry (see Fig.~\ref{OUTPLANE_electricfield}(b-c)). 
We noticed that the band gap decreases successively with the electric field and becomes zero at a critical electric force field qE$_c$=0.077 eV/{\AA} as shown in Fig.~\ref{OUTPLANE_electricfield}(b). A further increase in the electric field opens up the band gap again as shown in Fig.~\ref{OUTPLANE_electricfield}(c), leading to the trivial insulating phase. This phenomenon is quite similar to TMDs. Furthermore, an analysis based on $\mathbb Z_2$ invariant and edge-state dispersion (see ~\cref{OUTPLANE_electricfield}) shows that the band closing drives a topological phase transition to a trivial state with $\mathbb Z_2=0$. This topological phase transition destroys the helical edge states by switching off the QSH state of the 1T$^\prime$-MoGe$_2$Z$_4$. Such a tunability of the charge/spin conductance of the helical edge state can be used as electrical control of an ON/OFF switch which could lead to a QSH-based device.

Since the crystal symmetries of the 1T$^\prime$-MGe$_2$Z$_4$ materials are the same as the 1T$^\prime$-WTe$_2$, the devices conceived for the WTe$_2$ can be proposed for the MGe$_2$Z$_4$ materials too with the advantage of a larger inverted gap. 
One possible application is the realization of the topological transistor where the interface with a wide gap insulator bearing small lattice mismatch is needed to protect the helical edge channels from being gapped by the interlayer hybridization\cite{Qian2014Quantum}.
Due to the difference in lattice constant respect to WTe$_2$, we propose h-BN and CoBr$_2$ as the suitable wide gap insulators, which bear comparable lattice constants.

{\it Conclusion:}
In summary, by means of first-principles calculations, we predict a new thermodynamically stable QSH insulating phase in the 1T$^\prime$ crystal structure of the 2D materials MGe$_2$Z$_4$ family (M= Mo or W, and Z= P or As). Our results indicate that the 1T$^\prime$-phase can be stabilized in the MGe$_2$Z$_4$ family, introducing the band inversion at $\Gamma$ with consequent topological phases. A large band gap of ~237 meV is found, interestingly, that is the largest inverted gap among the existing 1T$^\prime$-phases of TMDs and MSi$_2$Z$_4$ hosting QSH phases. The large spin Berry curvature around the spin-orbit induced anticrossing gap with an inverted band gap leading to a quantized SHC. The MoGe$_2$P$_4$ compound shows a  perfect quantization of SHC in the band gap with a value of 1.0$\frac{e^2}{h}$. We have illustrated the electric field driven topological phase transition i.e QSH insulator to a trivial insulator phase with Rashba-like splitting.
Our results suggest that fast switching from the QSH state to the trivial one can be achieved in the MoGe$_2$P$_4$ at an electric field of ~0.077 eV/{\AA}. 
The faster switching with respect to the MoSi$_2$P$_4$ compound is attributed to a weaker electronegativity of the Ge atoms on the external layers, therefore, the electric field acts on the Mo-states more efficiently. 
The appearance of a large band gap and fast switching from non-trivial to trivial phase in this class of materials could be very promising in designing room-temperature topological field-effect transistors..\\

{\it Acknowledgement:} We acknowledge M. S. Bahramy and Barun Ghosh for useful discussions.
The work is supported by the Foundation for Polish Science through the International Research Agendas program co-financed by the European Union within the Smart Growth Operational Programme and  the National Science Center in the framework of the "PRELUDIUM" (Decision No.: DEC-2020/37/N/ST3/02338).  C. A. acknowledges support by Narodowe Centrum Nauki (NCN, National Science Centre, Poland) Project No. 2019/34/E/ST3
/00404. Z.M acknowledge National Natural Science Foundation of China Grant. No. 62150410438.
We acknowledge the access to the computing facilities of the Interdisciplinary
Center of Modeling at the University of Warsaw, Grant G84-0, GB84-1,  g87-1117 and GB84-7.
We acknowledge the access to the computing facilities of the Poznan Supercomputing and Networking Center Grant No. 609.

\bibliography{main_BSversion.bib}
\end{document}